\documentclass[12pt]{article}

\usepackage{amssymb}
\usepackage{amsmath}
\usepackage{amsthm}
\usepackage{amsfonts}
\usepackage{color,soul}
\usepackage{array}
\usepackage{dsfont}
\usepackage{epsfig}
\usepackage{multicol}
\usepackage{caption}
\usepackage{psfrag}
\usepackage{xcolor}
\usepackage{graphicx}
\usepackage{authblk}
\input epsf

\numberwithin{equation}{section}

\title{\bf The Critical Domain Size of Stochastic Population Models}
\author[1]{Jody R. Reimer}
\author[2]{Michael B. Bonsall}
\author[1,3]{Philip K. Maini}
\affil[1]{Mathematical and Statistical Sciences, 632 Central Academic Building, University of Alberta, Edmonton AB, T6G 2G1, Canada}
\affil[2]{Mathematical Ecology Research Group, Department of Zoology, University of Oxford, Tinbergen Building, South Parks Road, Oxford, OX1 3PS, UK}
\affil[3]{The Wolfson Centre for Mathematical Biology, Mathematical Institute, University of Oxford, Andrew Wiles Building, Radcliffe Observatory Quarter, Woodstock Road, Oxford, OX2 6GG, UK}

\begin{document}
\maketitle

\maketitle

\begin{abstract}
Identifying the critical domain size necessary for a population to persist is an important question in ecology. Both demographic and environmental stochasticity impact a population's ability to persist. Here we explore ways of including this variability. We study populations which have traditionally been modelled using a deterministic integrodifference equation (IDE) framework, with distinct dispersal and sedentary stages. Individual based models (IBMs) are the most intuitive stochastic analogues to IDEs but yield few analytic insights. We explore two alternate approaches; one is a scaling up to the population level using the Central Limit Theorem, and the other a variation on both Galton-Watson branching processes and branching processes in random environments. These branching process models closely approximate the IBM and yield insight into the factors determining the critical domain size for a given population subject to stochasticity.
\end{abstract}


\section{Introduction}

Determining the size of the domain necessary for population persistence is a well known problem in ecology. First discussed in \cite{Kierstead1953}, the \textit{critical domain size} is the domain size required to independently sustain a population. For deterministic models, the critical domain size may be described as the smallest domain for which there exists a stable non-zero steady state. However, at low population densities, deterministic models may not accurately reflect extinction risks. For example, when the need for a reserve or management area arises, the population of interest is often already depleted and thus more vulnerable to variability both in demographic rates and in its environment. For populations subject to uncertainty, we broadly define the critical domain size as the size of the domain necessary to reach some accepted measure of persistence. One such measure is the \textit{probability of extinction} of a population, that is, the probability of the total population size being zero by a given time. What is deemed to be an acceptable probability of extinction will vary depending on application (e.g. \cite{Burnham2002, Flather2011}). In this work, we will use the \textit{probability of ultimate extinction}, which is the limit of the probability of extinction as time tends to infinity. The simplest domain, and the one we consider in this work, is a one-dimensional domain of length $L$, which is both a good starting point to develop theory as well as biologically reasonable in certain habitats such as rivers, shorelines, or narrow valleys. We will also assume very harsh conditions outside of the domain (e.g. strong negative anthropogenic influence, uninhabitable landscapes) where no individuals survive. Thus the critical domain size estimated here may be conservative if individuals may survive outside of the domain.

\paragraph{Integrodifference equations}

We are interested in populations with discrete dispersal and reproductive phases, commonly modeled using integrodifference equations (IDEs). Many plants, for example, disperse as seeds for a short time and then remain in one place for the remainder of their lives, as do several marine species. For simplicity, we consider only non-structured populations (i.e. all of the individuals disperse and then reproduce) in order to be able to determine the effects of stochasticity on the extinction dynamics without conflating the results with those caused by more complicated population structures. We assume the simplest case of density independent linear population growth in order to separate the effects of stochasticity from those due to any nonlinearities. We also note that many of the populations we are interested in will be well below the environmental carrying capacity and a linear growth rate is a close approximation to many nonlinear growth functions at very low densities in the absence of an Allee effect. We here model only females, assuming males are sufficient for reproduction.

An IDE satisfying the above assumptions has the form 
\begin{equation}
N_{n}(x) = \begin{cases} \int_{-L/2}^{L/2}{k(x,y)r\,N_{n-1}(y)dy}, & x \in \left[-L/2, L/2\right]\\
 0, & x < -L/2 \mbox{ or } x > L/2 \end{cases}
\label{eq:IDEform}
\end{equation}
where $N_{n}(x)$ is the population density in generation $n$ at location $x \in \mathbb{R}$, 
$r$ denotes the linear population growth rate and $k(x,y)$ is the dispersal kernel, a probability density function describing the probability of movement of an individual during one time step from location $y$ to $x$. For example, the Laplace dispersal kernel, 
\begin{equation}
k(x,y) = \sqrt{\frac{\alpha}{4D}}\exp{\left( -|x-y|\sqrt{\frac{\alpha}{D}}\right)}  
\label{eq:Laplace}
\end{equation}
arises from the assumption that organisms settle out of a pelagic dispersal phase (diffusion coefficient $D$) at a constant rate $\alpha$ \cite{Lutscher2005}. As has been done previously \cite{Reimer2015}, we scale the dispersal kernels using the mean of the strictly positive distribution (i.e. the mean of $k(x)$ for $k \geq 0$) and will refer to this as the mean dispersal distance. The Laplace kernel has a mean value of $\sqrt{D/\alpha}$ \cite{Kot1992, Lockwood2002} and here all units are scaled to this mean dispersal distance for simplicity. We will use the Laplace kernel in the following examples, as it is one of the few kernels for which an analytic solution to the critical domain size problem for this deterministic IDE exists \cite{Reimer2015, VanKirk1997}. 

\paragraph{Existing theory}

The methods we will consider here have been motivated by, and applied widely to, studies of genetics (see e.g. \cite{Feller1951, Keiding1975}). To the best of our knowledge, stochastic analogues to integrodifference equations have only been studied for calculating invasion speeds of travelling waves subject to variability. Kot et al.\ \cite{Kot2004} used density independent branching random walks to simulate a stochastic IDE with linear, density independent growth, determining that stochastic variation in dispersal and reproduction does not lower the asymptotic invasion speed \cite{Kot2004}. We also build on previous results on non-spatial stochastic population models (e.g. \cite{Engen1998, Lande1993, Leigh1981, Hiebeler1997}). In this work, we combine these results for non-spatial stochastic models with stochastic analogues to IDEs in order to understand the effect of domain size on persistence in single species spatial stochastic models.

\paragraph{Stochastic models} We are interested in stochastic models analogous to IDEs for populations with discrete dispersal and life history events. In Section \ref{sec:demstoch}, we address the effects of demographic stochasticity, which affects the fitness of each individual in a population independently and arises from the variability inherent in deaths, dispersal, and reproductive events. We incorporate demographic stochasticity using random variables with given distributions ideally obtained through statistical study \cite{Engen1998, Lande1993, Shaffer1981}. Demographic stochasticity is most important for populations with low numbers, as individual level fluctuations in growth and death rates most significantly affect subsequent generations for smaller populations.

We first develop a spatially explicit individual-based model (IBM) which is the most realistic way to incorporate uncertainty into an IDE framework. In order to obtain more general analytic results, we develop two spatially implicit approximations to the IBM. The first scales the model up to the population level using the Central Limit Theorem. The second approximation allows for analytic results and is based on a Galton-Watson branching process modified to include loss via dispersal outside of the domain. We use classical results on branching processes to determine the domain size necessary to achieve specified conservation goals. Both of these approximations make use of the modified dispersal success approximation of Reimer et al.\ \cite{Reimer2015} which is based on the idea that we can avoid explicit spatial dependence by assuming that during each time step a fixed proportion of the population is lost via dispersal to areas outside the domain. We compare these results with both the corresponding deterministic IDE as well as simulations from the stochastic IBM through an illustrative example, though the results hold over a much broader parameter space.  

In Section \ref{sec:Env_stoch}, we consider environmental stochasticity which is important to populations of any size \cite{Lande1993}. This stochasticity arises from fluctuations in the environment that are assumed to affect the demographic rates of all individuals in a population simultaneously and in a similar way (e.g. variable temperatures or changes in food availability) \cite{Lande1993, Leigh1981}. We address the same question of critical domain size using similar methods but now allowing for environmental stochasticity as well. We reformulate our IBM to include dependence on a random environmental variable and again consider the dynamics approximated at the population level. The branching process framework used to investigate the effects of demographic stochasticity is modified to include environmental variation using the theory of branching processes in random environments. Both demographic and environmental variability are known to be important in natural populations \cite{Lande1988, Melbourne2008} and the tools developed here allow for further exploration of their comparative effects on extinction risk. 

\section{Demographic stochasticity} \label{sec:demstoch}
\subsection{Individual based models with demographic stochasticity} \label{sec:IBMs}

Simulating each individual in a population via an IBM is one of the simplest and most intuitive ways to examine population dynamics. Unfortunately, results generated in this way are difficult to compare and analytic tools have not yet been developed to determine the probability of extinction, making it difficult to determine sensitivity to various factors such as initial population size or demographic rates. We describe a stochastic IBM analogous to an IDE in order to be able to compare our approximations to these simulation results. 

Simulations begin at generation $n = 0$ with an initial population of size $N_0$ evenly distributed throughout a one-dimensional domain of length $L$. We consider only females, so each individual has $r$ offspring where $r$ is a random variable with probability mass function $\{p_i\}$ for $i = 0, 1, 2, \ldots$, where $p_i$ is the probability of an individual having $i$ female offspring in one reproductive period and $\sum{p_i} = 1$. Note that this process will not explicitly incorporate individual deaths. Hence $p_1$ is the probability that either the female produced one offspring and then died, or that she produced no offspring but survived to the next generation, and $p_0$ is the probability that an individual produced no offspring and did not survive until the next generation. Thus parental survival is included in the term ``offspring''. At the start of each generation, and for each individual, we first determine the random variable for the number of offspring produced, as drawn from $\{p_i\}$. Each of these offspring then disperses according to a dispersal kernel $k(x,y)$ and if they remain within the domain, they are assumed to survive to the start of the next time step and the process is repeated.

In order to investigate the probability of extinction by a given generation or the ultimate probability of extinction, it is necessary to run many simulations of this model. 

\subsection{Spatially implicit population level model}\label{sec:pop_level}

As has been done for IDE models, we can approximate the population level outcomes of the IBM by disregarding the location of individuals inside the domain and only considering changes to the total population size via a spatially implicit approximation. We do this by approximating the proportion of the population which successfully remains within the domain following the dispersal process from time $n-1$ to $n$ with a scalar $\bar{A}$. Note that throughout this work, a bar is used to denote population level variables. 

For the IDE model Eq.\ \eqref{eq:IDEform}, the total population size within the domain $\bar{N}_n$ at time $n$ has been approximated via the deterministic recursion relation
\begin{equation}
\bar{N}_{n} = \int_{-L/2}^{L/2}{\int_{-L/2}^{L/2}{k(x,y)r\,N_{n-1}(y)dy}dx} \approx \bar{A}r\bar{N}_{n-1}
\label{eq:approximation_eqn}
\end{equation}
%
where $0 \leq \bar{A} \leq 1$ \cite{Reimer2015, VanKirk1997}. We apply this same concept when approximating the total population size of the stochastic IBM according to
\begin{equation}
\bar{N}_n = \bar{A}\bar{r}\bar{N}_{n-1}.
\label{eq:approximation_IBM_eqn}
\end{equation} 
The IBM assumes that the number of offspring of each individual in a population is a real-valued independent identically distributed (iid) random variable $r$ with probability mass function $\{p_i\}$, expected value $\mathbb{E}[r] = \xi$, and variance $\sigma_r^2$. To scale from the individual to the population level, we employ the Central Limit Theorem; for a population of size $\bar{N}_n$, the average number of offspring per capita $\bar{r}_{\bar{N}}$ converges to a normally distributed random variable with mean $\mathbb{E}[\bar{r}_{\bar{N}}] = \xi$ and variance $\mbox{Var}(\bar{r}_{\bar{N}}) = \sigma_r^2/\bar{N}_n$. 
Note $\bar{r}_{\bar{N}}$ cannot realistically take a negative value, so we assume the distribution of $\bar{r}_{\bar{N}}$ to be normally distributed as above but with all of the probabilities of a negative distribution assigned to zero and the distribution then appropriately rescaled.

In order to remove explicit spatial dependence of each individual, we consider a scalar $A$ which is the probability that an individual remains within the domain following one dispersal period. We first naively assume individuals are evenly distributed throughout a one-dimensional domain of length $L$ at any given time. For an individual at position $y \in \left[-L/2, L/2\right]$, this individual settles within the domain with probability $s(y)$ \cite{VanKirk1997}, where
\begin{equation}
s(y) = \int_{-L/2}^{L/2}{k(x,y)dx}.
\end{equation}
By the Central Limit Theorem, the proportion $\bar{A}_{\bar{N}}$ of the population of size $\bar{N}_n$ which remains within the domain is a normally distributed random variable. For individuals assumed to be evenly distributed throughout the domain, $\mathbb{E}[\bar{A}_{\bar{N}}] = S$, where $S$ is the average dispersal success \cite{VanKirk1997}:
\begin{equation}
S = \frac{1}{L}\int_{-L/2}^{L/2}{s(y)dy}.
\end{equation}
The variance of the distribution of $\bar{A}_{\bar{N}}$ is $\mbox{Var}(\bar{A}_{\bar{N}}) = \sigma_S^2/\bar{N}_n$, where $\sigma_S^2$ is the variance of $s(y)$ over the domain:
\begin{equation}
\sigma_S^2 = \mbox{Var}(s(y)) = \int_{-L/2}^{L/2}{(s(y) - S)^2dy}.
\end{equation}
While $S$ provides a reasonable approximation to the average dispersal success of a population for a range of deterministic IDEs (this was explored in \cite{Reimer2015}), our recent work has revisited the assumption that the population is evenly distributed throughout the domain. We use the modified dispersal success approximation $\widehat{S}$ throughout this work, which we found in \cite{Reimer2015} consistently outperforms $S$ as an approximation to the proportion of individuals retained in the domain following dispersal. To obtain $\widehat{S}$, we weight the dispersal success functions by $s(y)/S$ when approximating the population's average successful dispersal rate. This is based on the assumption that more individuals are in the centre of the domain than near the edges and that $s(y)$ provides a good approximation to the population distribution inside the domain \cite{Reimer2015, VanKirk1997}. 
This modified average dispersal rate is:
\begin{equation}
\widehat{S} = \frac{1}{L}\int_{-L/2}^{L/2}{\left(\frac{s(y)}{S}\right)s(y)dy}.
\label{eq:hatS}
\end{equation} 
It has been shown for IDEs that $\widehat{S} \geq S$ \cite{Reimer2015}, since it weights areas which are thought to have more individuals more heavily, based on the idea that they not only have higher dispersal success but are also at a location that is more likely to receive settling individuals for symmetric dispersal kernels. At the population level, $\bar{A}_{\bar{N}}$ may be drawn from a normal distribution with $\mathbb{E}[\bar{A}_{\bar{N}}] = \widehat{S}$ and variance $\mbox{Var}(\bar{A}_{\bar{N}}) = \sigma_{\widehat{S}}^2/\bar{N}_n$, where $\sigma_{\widehat{S}}^2$ is the variance of $s(y)^2/S$ over the domain: 
\begin{equation}
\sigma_{\widehat{S}}^2 = \mbox{Var}(s(y)^2/S) = \int_{-L/2}^{L/2}{\left(\frac{s(y)^2}{S} - \widehat{S}\right)^2dy}.
\label{eq:hatSvariance}
\end{equation}
We call this the $\widehat{S}$ population level approximation for $\bar{A}_{\bar{N}}$. Now in each generation we start with a population of size $\bar{N}_{n-1}$, determine the values of the random variables $\bar{r}_{\bar{N}}$ and $\bar{A}_{\bar{N}}$ according to their distributions as described above, and calculate $\bar{N}_{n}$ according to Eq.\ \eqref{eq:approximation_IBM_eqn}. 

It is important to note here that we now redefine our notion of extinction from requiring the death of all individuals to requiring the population size to fall below a certain value. This is to account for the fact that even if $\bar{A}_{\bar{N}}$ is less than 1 for many generations, the population level approximation will only tend asymptotically towards zero but not achieve it. This threshold, which we will call the quasi-extinction threshold (e.g. \cite{Ginzburg1982, Lande1993}), will here be $1$, so that if $\bar{N} < 1$, our population will be considered extinct. This threshold value of $1$ will be used for all population level approximations in this work. 

\subsubsection{A brief note on the time to extinction} \label{sec:time_ext}

The number of surviving offspring of an individual over a time step is determined by the product of the two positive random variables $r$ and $A$. We could also consider $B = Ar$ as a single independent random variable. 
Since $A$ and $r$ are independent,  
\begin{equation}
\mathbb{E}[B] = \mathbb{E}[A]\,\mathbb{E}[r]
\end{equation}
and 
\begin{equation}
\mbox{Var}(B) = \sigma_B^2 = \mathbb{E}[A]^2\, \sigma_r^2 + \mathbb{E}[r]^2\, \sigma_A^2 + \sigma_A^2\, \sigma_r^2.
\end{equation}
Since we are here interested in the population level model, we can ignore the shape of the distribution of $B$ and look to its average value. For a population of size $\bar{N}_n$, the population level growth rate over one generation is $\bar{B} \sim \mathcal{N}(\mathbb{E}[B],\sigma_B^2/\bar{N}_n)$ by the Central Limit Theorem. 

\begin{figure}[!h]
\begin{center}
\includegraphics[width=0.6\textwidth]{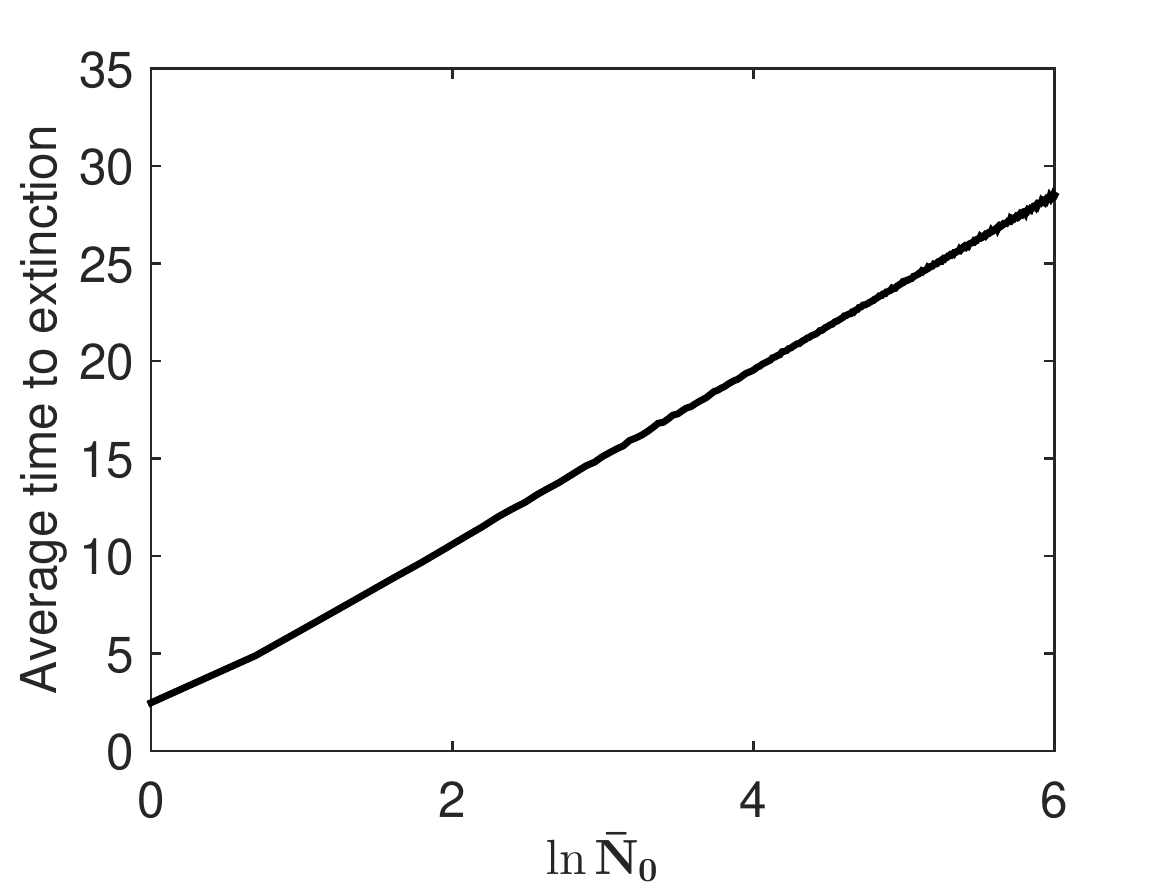}
\caption{Average time to extinction of Eq.\ \eqref{eq:Bapprox} with varying initial conditions $\bar{N}_0$ for a population with certain ultimate extinction ($\mathbb{E}[\bar{B}] < 1$) as described in Section \ref{sec:time_ext}. Here $\mathbb{E}[\bar{B}] = 0.8$ and $\mbox{Var}(\bar{B}) = 0.3$. As predicted by the deterministic model as well as the continuous time stochastic model of \cite{Lande1993}, the average time to extinction scales linearly with $\ln{\bar{N}_0}$. This plot was made from 10000 simulations for each value of $\bar{N}_0$ with a quasi-extinction threshold of $\bar{N}_n < 1$. All simulations in this work were performed using commercial software packages (MATLAB and Statistics Toolbox Release R2014b, The MathWorks, Inc., Natick, Massachusetts, United States)}\label{fig:time_to_ext_pop_level_scaling}
\end{center}
\end{figure}

For a deterministic model of the form $\bar{N}_{n+1} = \bar{B}\, \bar{N}_n$ where $\bar{B}$ is a constant scalar, if $\bar{B} > 1$, the population tends to infinity and for $\bar{B} < 1$, the population tends to extinction. For the deterministic model with $\bar{B} < 1$ and an initial population of size $\bar{N}_0$, we can calculate the time to extinction by setting $\bar{B}^n \bar{N}_0 = 1$ (recall that the quasi-extinction threshold is 1) and solving for $n$ as 
\begin{equation}
n = \frac{-\ln{(\bar{N}_0)}}{\ln{(\bar{B})}}.
\end{equation}
In the stochastic population level model 
\begin{equation}
\bar{N}_{n+1} = \bar{B}\, \bar{N}_n,
\label{eq:Bapprox}
\end{equation}
where $\bar{B}$ is a random variable as described above, the expected value of $\bar{N}_{n}$ is $\bar{B}^{n}\bar{N}_0$, but now even with $\bar{B} > 1$, extinction may still occur. As in the deterministic model, linear dependence of $n$ on $\ln(\bar{N}_0)$ was found by Lande and Russell \cite{Lande1993} when considering the time to extinction of a continuous time stochastic model. This dependence on $\ln(\bar{N}_0)$ seems to also hold here when we include space implicitly into our discrete time stochastic model (Figure \ref{fig:time_to_ext_pop_level_scaling}).

\subsubsection{Example of the population level approximation}\label{sec:pop_level_demo}
\begin{figure}[!h]
	\begin{center}
		\includegraphics[width=1.0\textwidth]{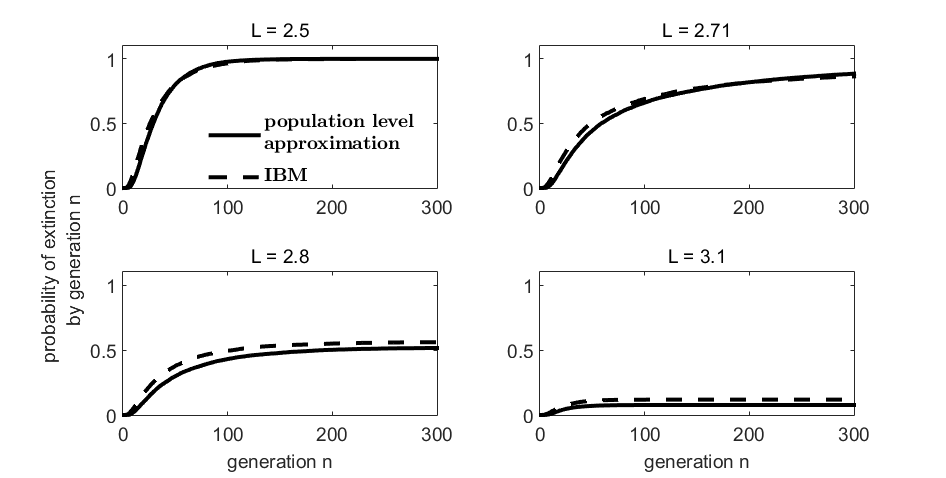}
	\caption{The stochastic population level approximation to the IBM according to Eq.\ \eqref{eq:approximation_IBM_eqn} for the example of Section \ref{sec:pop_level_demo} closely approximates the cumulative extinction probabilities of the IBM.
	10000 simulations were run from an initial population $\bar{N}_0 = 10$. 
	The domain values represented are around the critical domain size $L^* = 2.703$ of the corresponding deterministic IDE model with $r = 1.5$, logistic growth, and a Laplace dispersal kernel as described in \cite{Reimer2015} 
	}
	\label{fig:population_level_stochastics}
	\end{center}
\end{figure}

Let offspring be produced according to the probability mass function $\{p_0 = 0.1,\, p_1 = 0.3,\, p_2 = 0.6\}$, so that $\mathbb{E}[\bar{r}_{\bar{N}}] = 1.5$ and $\mbox{Var}(\bar{r}_{\bar{N}}) = 0.45/\bar{N}_n$, and suppose dispersal follows a Laplace distribution Eq.\ \eqref{eq:Laplace}.  If we assume that the average dispersal success approximation $\widehat{S}$ describes the loss of individuals in one generation outside the domain of length $L$, then $\bar{A}_{\bar{N}}$ is a random variable chosen from a normal distribution with mean 
\begin{equation}
\mathbb{E}[\bar{A}_{\bar{N}}] = \widehat{S} = \frac{2\, L\, \mathrm{e}^{\frac{L}{b}} - b + 4\, L\, \mathrm{e}^{\frac{2\, L}{b}} + 8\, b\, \mathrm{e}^{\frac{L}{b}} - 7\, b\, \mathrm{e}^{\frac{2\, L}{b}}}{4\, \mathrm{e}^{\frac{L}{b}}\, \left(b + L\, \mathrm{e}^{\frac{L}{b}} - b\, \mathrm{e}^{\frac{L}{b}}\right)}
\label{eq:Shat_Lap}
\end{equation}
and variance
\begin{align*}
\mbox{Var}(\bar{A}_{\bar{N}}) = \frac{\sigma_{\widehat{S}}^2}{\bar{N}_n} &= \Big(\frac{7}{2} + \left(S^2 - 2 + S^{-2} - \mathrm{e}^{-L}\right)L  - \frac{269}{96\, S^2} - \frac{4}{\mathrm{e}^{L}} \\ 
& + \frac{1}{2\, \mathrm{e}^{2\, L}} + \frac{3\, L}{S^2\, \mathrm{e}^{L}} + \frac{3\, L}{8\, S^2\, \mathrm{e}^{2\, L}} + \frac{5}{4\, S^2\, \mathrm{e}^{L}} + \frac{3}{2\, S^2\, \mathrm{e}^{2\, L}} \\
& + \frac{1}{12\, S^2\, \mathrm{e}^{3\, L}} - \frac{1}{32\, S^2\, \mathrm{e}^{4\, L}}\Big)\Big{/}\bar{N}_n
\end{align*}

according to Eq.\ \eqref{eq:hatS} and Eq.\ \eqref{eq:hatSvariance} respectively. This $\widehat{S}$ population level approximation closely predicts the cumulative extinction probabilities (the probability of extinction by a given generation) of the IBM (Figure \ref{fig:population_level_stochastics}).

\subsection{Modified Galton-Watson branching process}\label{sec:Branching}

While the population level simulations provide a reasonable and less computationally costly approximation to the IBM, they do not provide any new understanding of the driving processes or variables. We turn to a different approximation in order to be able to explore extinction probabilities and mechanisms analytically. We consider a modified Galton-Watson process adapted to suit our populations of interest by including an extra spatially implicit dispersal step following the reproduction process. Using branching processes allows us to deal with a population of discrete individuals experiencing stochastic variation in both their dispersal and reproduction \cite{Kot2004}.

\subsubsection{Galton-Watson branching processes} 

Originally motivated by the desire of the French aristocracy to predict the extinction of family names \cite{Keiding1975, Watson1875}, Galton-Watson branching processes have since been applied to studies of nuclear chain reactions, the survival of mutant genes, and population biology \cite{Feller1951}. They are based on the assumption that individuals in each generation have $r$ offspring, independently of each other, where $r$ is an iid random variable with probability mass function $\{p_i\}$, $i = 0, 1, 2, \ldots$. Each $p_i$ is the probability that an individual produces $i$ offspring in one generation, so $\sum{p_i} = 1.$ This corresponds to the probability generating function $f(s) = \sum p_i s^i$ for $s$ on the unit interval. 

The population size is a sequence of random variables $\{ Z_n \}$ describing the population size in the $n^{th}$ generation, given an initial population $Z_0 = z$. The mean reproductive rate is
\begin{equation}
\xi = \mathbb{E}[r] =  \sum_{i=0}^\infty{i\,p_i} = f'(1),
\end{equation}
and the expected size of the $n^{th}$ generation is 
\begin{equation}
\mathbb{E}[Z_n] = z\,\xi^n.
\end{equation}
Unlike in similar deterministic models, there remains a chance of extinction even when the mean reproductive rate is greater than 1. The realized population size $Z_{n+1}$ is found via a recursion relation which sums the descendants of each of the individuals in generation $n$, so that
\begin{equation}
Z_{n+1} = \sum_{j = 0}^{Z_n}{r_{j,n}}
\end{equation}
where $r_{j,n}$ is an integer-valued random variable describing the number of offspring of the $j$th individual in generation $n$  \cite{Watson1875}. 

To understand the classical results on the extinction probabilities for Galton-Watson branching processes, we first consider the case where $Z_0 = 1$, since the results for $Z_0 = z > 1$ follow easily. The probability $d_n$ of extinction by time $n$ of the lineage of one individual in the first generation can be found via the intuitive recursion relation
\begin{equation}
d_n = \underbrace{p_0}_{(\mbox{a})} + \underbrace{p_1 d_{n-1}}_{(\mbox{b})} + \underbrace{p_2 (d_{n-1})^2}_{(\mbox{c})} + \ldots. 
\label{eq:standard_ext_eq}
\end{equation}
Here (a) is the probability that the original individual had no offspring, (b) is the probability that it produced one offspring and that this individual's lineage went extinct in the next $n-1$ generations, and similarly, (c) is the probability that it produced two offspring and that both of their lineages died out in the subsequent $n-1$ generations, with the pattern continuing for all possible numbers of offspring \cite{Bartlett1960}.

The ultimate extinction probability $d$ is the limit
\begin{equation}
d = \lim_{n\rightarrow\infty}{\Pr{(Z_n = 0)}}.
\end{equation}
Clearly $d_n$ must reach a limit as $n$ increases, since it is an increasing function in $n$ and is bounded above by 1, so as it reaches this limit, $d_{n-1}$ tends to $d_{n}$. We thus find the asymptotic extinction probability by setting $d_n = d_{n-1} = d$ in Eq.\ \eqref{eq:standard_ext_eq} and solving
\begin{align}
d &= p_0 + p_1 d + p_2 d^2 + p_3 d^3 + \ldots \nonumber \\
  &= \sum_{i=0}^\infty{p_i\, d^i} = f(d)
\label{eq:nonspatial_ult_ext}
\end{align}
%
where $f(d)$ is the probability generating function in $d$. This always has a solution of $d = 1$, but may also possess further solutions less than 1. A trivial case occurs when each individual produces exactly one individual in the subsequent generation, in which case $p_1 = 1$ and the probability of extinction is $d = 0$ so the population size remains constant over time. 
Disregarding the trivial case, Galton and Watson's classic results show that if the mean number of offspring produced by an individual $\xi$ (recall that $\xi = f'(1)$) is less than or equal to one (known as the \textit{subcritical} and \textit{critical} cases, respectively), then the population will die out almost surely. If $\xi$ is larger than one (the \textit{supercritical} case), then the extinction probability is strictly less than 1 and the population will tend to grow exponentially \cite{Bartlett1960, Keiding1975}.

\subsubsection{A spatially implicit modified branching process}\label{sec:spatial_bp}

We consider the possibility that, in addition to the birth-and-death processes of the Galton-Watson model, a certain fraction $F$ of each new generation is lost via dispersal outside the domain where they die before they can reproduce. Now the probability of extinction by generation $n$ is dependent upon both the usual birth-death processes as well as the dispersal behaviour of the individuals. We maintain consistency with the IBM regarding the order of these two processes within one generation so that individuals first reproduce according to the probability generating function $f(s)$ and then disperse, leaving the domain with probability $F$. We thus obtain a modified version of Eq.\ \eqref{eq:standard_ext_eq},
\begin{gather}
\begin{aligned}
d_n = \underbrace{p_0}_{(\mbox{a})} &+ \underbrace{p_1\, d_{n-1}(1-F)}_{(\mbox{b}_1)} + \underbrace{p_1\, F}_{(\mbox{b}_2)} + \underbrace{p_2\, (d_{n-1}(1-F))^2}_{(\mbox{c}_1)} \\
	& + \underbrace{p_2\, F^2}_{(\mbox{c}_2)} + \underbrace{2\, p_2\, F(1-F)\, d_{n-1}}_{(\mbox{c}_3)}\ldots. 
\end{aligned}
\label{eq:ext_eq}
\end{gather}
%
%
Here (a) is the probability that the original individual had no offspring, (b$_1$) describes the probability that the first individual produced one surviving offspring and that this individual's lineage went extinct in the subsequent $n-1$ generations, and (b$_2$) is the probability that the single offspring of the first individual dispersed to unfavorable habitat before it could reproduce in the next generation. Following intuitively, (c$_1$) is the probability that the original individual produced two offspring and both of their lineages died out in $n-1$ generations, (c$_2$) is the probability that the two offspring dispersed to unfavourable habitat and were lost, and (c$_3$) is the probability that only one of them settled outside the domain while the other remained but their lineage died out in $n-1$ generations, and the recursion continues on in this way. 

We may re-write Eq.\ \eqref{eq:ext_eq} as
\begin{equation}
d_n = \sum_{v=1}^{\infty}{\sum_{q=1}^{v}{{v \choose q} p_{v-1}\, [(1-F)d_{n-1}]^{q-1}\,F^{(v-q)}}} 
\end{equation}
%
and so the probability of extinction as $n \rightarrow \infty$ is found by solving for $d$ in 
\begin{align}
d &= \sum_{v=1}^{\infty}{\sum_{q=1}^{v}{{v \choose q} p_{v-1}\, [(1-F)d]^{q-1}\,F^{(v-q)}}}\nonumber \\
  &= f[(1-F)d + F] \label{eq:prob_ext_disp}
\end{align}
where $f[(1-F)d + F]$ is the probability generating function in $(1-F)d + F$. This is seen by rearranging Eq.\ \eqref{eq:prob_ext_disp} as 
\begin{equation}
d = p_0 + p_1 ((1-F)d + F) + p_2 ((1-F)d + F)^2 + p_3 ((1-F)d + F)^3 + \ldots. 
\end{equation}
Trivially, if there is no good habitat, the probability of extinction by any time step will be 1, since $d_n = p_0 + p_1 + p_2 + \ldots$ when $F = 1$ and so $d = 1$ is always a solution to Eq.\ \eqref{eq:prob_ext_disp}. Whether there is a second, smaller solution is now determined not only by the reproductive probabilities but also by $F$.

To determine whether the process is supercritical, critical, or subcritical and thus determine whether extinction is certain we need to determine the expected value of offspring of this modified branching process. Define $\tilde{p}_i$ as the probability of an individual producing $i$ offspring which successfully settle within the domain following their dispersal phase. Then

\begin{align}
\tilde{p}_0 &= p_0 + p_1F + p_2F^2 + \ldots \nonumber \\ 
\tilde{p}_1 &= p_1(1-F) + 2p_2(1-F)F + 3p_3(1-F)F^2 + \ldots \nonumber\\
\tilde{p}_2 &= p_2(1-F)^2 + 3p_3(1-F)^2F + 6p_4(1-F)^2F^2 + \ldots \nonumber\\
& \vdots \nonumber\\
\tilde{p}_q &= \sum_{v = q}^\infty{{v \choose q,v-q}p_v(1-F)^qF^{v-q}} \label{eq:coeff_eqn}
\end{align}
where
\begin{equation}
{v \choose q,v-q} = \frac{v!}{q!(v-q)!}
\end{equation}
are multinomial coefficients. The corresponding probability generating function is 
\begin{align}
f(s) = \tilde{p}_0 + \tilde{p}_1s + \tilde{p}_2s^2 + \tilde{p}_3s^3 + \ldots 
\end{align}
and so the expected reproductive value is 
\begin{align}
\xi &= f'(1) = \tilde{p}_1 + 2\tilde{p}_2 + 3\tilde{p}_3 + \ldots \nonumber\\
	&= \sum_{q=1}^{\infty}{q\tilde{p}_j} = \sum_{q=1}^{\infty}{\sum_{v = q}^\infty{q\,{v \choose q,v-q}p_v(1-F)^qF^{v-q}}}. \label{eq:expected_rep}
\end{align}

If we make the assumption that these sums do not run to infinity but rather to some maximum number of biologically possible offspring, then these summations become finite. Eq.\ \eqref{eq:expected_rep} simplifies to 
\begin{equation}
\xi = \mathbb{E}[r]\, (1-F),
\label{eq:ind_result}
\end{equation}
which can be shown by induction for both the infinite or finite sum (Appendix \ref{Appendix:induction}). Thus the expected reproductive rate of the modified branching process is the product of the expected reproductive rate without the loss of individuals outside the domain and the proportion of individuals which successfully settle within the domain. Applying the results of Watson and Galton \cite{Watson1875}, the criticality condition for the spatially implicit branching process is
\begin{equation}
\xi = \mathbb{E}[r]\,(1-F) = 1,
\label{eq:crit_cond_bp}
\end{equation}
and from this we can determine the proportion of individuals $F^*$ which must be retained within the domain in order to avoid certain eventual extinction. For any $F<F^*$, the population is supercritical and for any $F \geq F^*$, eventual extinction is certain. To find a suitable value for $F$, we recall the dispersal success approximation $\widehat{S}$ of Eq.\ \eqref{eq:hatS}. We let $F = 1-\widehat{S}$ and compare the resulting extinction risks with those from the spatially explicit IBM simulation.

\subsubsection{Example of modified branching process}\label{sec:ex_branching}

\begin{figure}[!h]
	\begin{center}
		\includegraphics[width=1.00\textwidth]{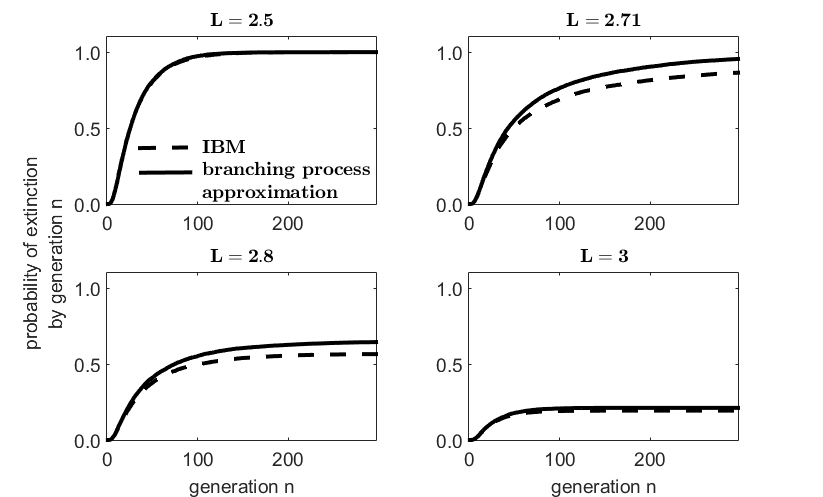}
	\caption{Modified branching process approximations remain close to the IBM simulation on domains of various sizes close to the critical values predicted by both the deterministic IDE model ($L^* = 2.703$) and the $\widehat{S}$ branching process approximation model ($L_{\widehat{S}}^* = 2.722$) as described in Section \ref{sec:ex_branching}. For each plot, 10000 simulations were performed on an initial population $z = 10$}
	\label{fig:spatial_stochastic_model_pop1p2}
	\end{center}
\end{figure}

Suppose an individual in a population of initial size $Z_0 = z$ reproduces according to $\left\{p_0 = 0.1, p_1 = 0.3, p_2 = 0.6\right\}$, for a mean reproductive rate $\mathbb{E}[r] = 1.5$. Assume dispersal occurs according to a Laplace distribution Eq.\ \eqref{eq:Laplace}. As in \cite{Reimer2015}, the critical domain size of the deterministic IDE model with logistic growth, an intrinsic growth rate of $r = 1.5$, and a Laplace dispersal kernel is $L^* = 2.703$ times the mean dispersal distance. For a domain of this size, $\widehat{S} = 0.6647$ (from Eq.\ \eqref{eq:Shat_Lap}) and so $\mathbb{E}[r]\,(1-F) = \mathbb{E}[r]\, \widehat{S} = 0.9970 < 1$. Thus the branching process predicts that the critical domain size of the deterministic model is too small to sustain a population subject to demographic stochasticity (Figure \ref{fig:spatial_stochastic_model_pop1p2}). 

\begin{figure}[!h]
	\begin{center}
		\includegraphics[width=0.70\textwidth]{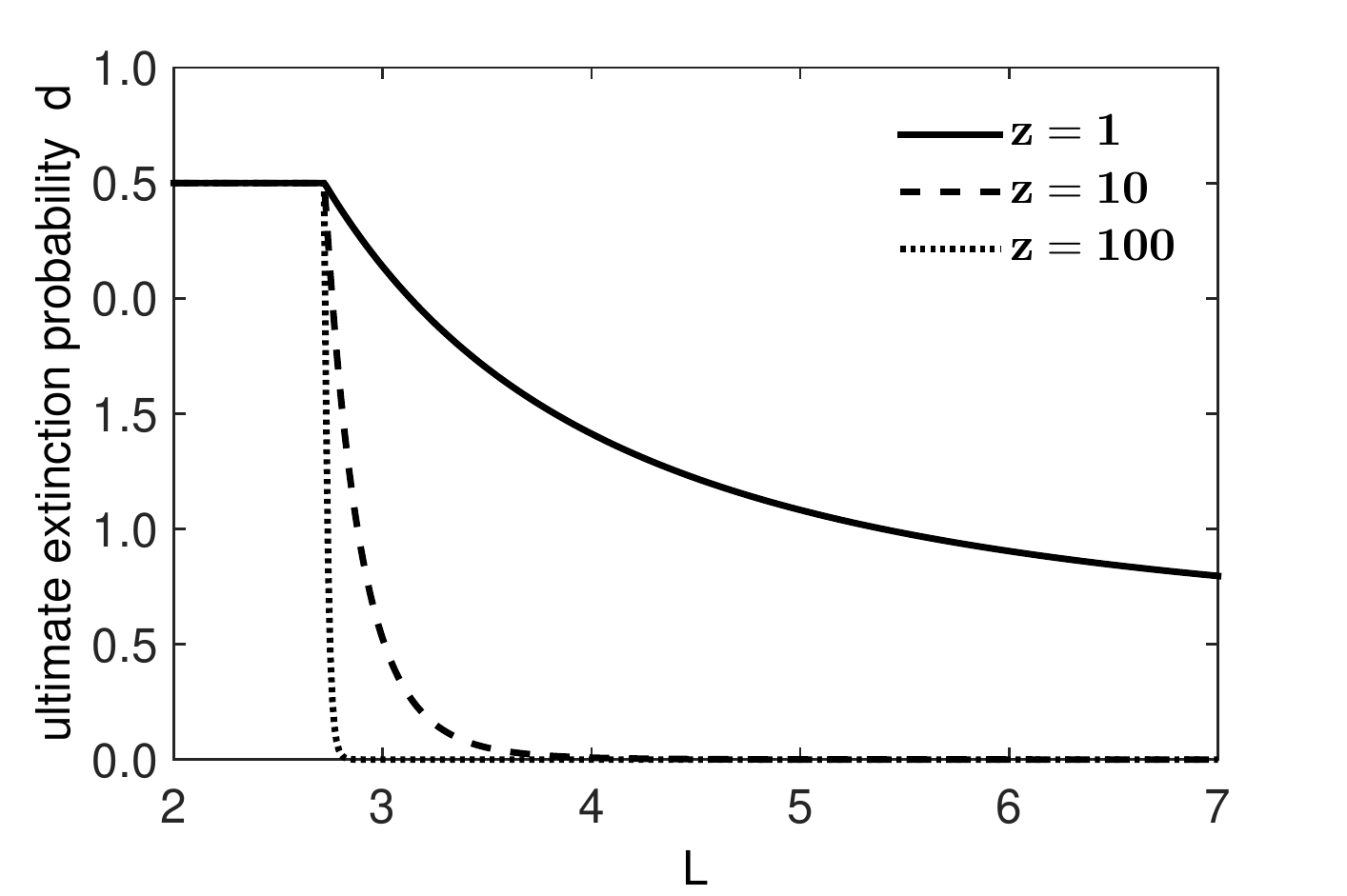}
	\caption{Probability of ultimate extinction for the $\widehat{S}$ branching process model dependent on domain length and initial population size $z$ for the example of Section \ref{sec:ex_branching}. For increasing initial values, the benefit of increasing the domain size $L$ much beyond the critical length in terms of the ultimate extinction probability is negligible  
	}
	\label{fig:prob_extinct_against_L}
	\end{center}
\end{figure}

We compare the probability of eventual extinction for a range of domain lengths around $L^*$ (Figure \ref{fig:prob_extinct_against_L}). The branching process model predicts certain extinction until some critical length value, at which point the ultimate extinction probability becomes less than one. The greater the initial population, the sharper the reduction in the ultimate extinction probability for lengths larger than the critical value. The critical domain length at which the population's ultimate extinction probability becomes less than one can be found by solving the criticality condition Eq.\ \eqref{eq:crit_cond_bp} for $(1-F) = \widehat{S} = 1/\mathbb{E}[r]$ and then solving for $L$ using Eq.\ \eqref{eq:Shat_Lap}, obtaining $L_{\widehat{S}}^* = 2.722$ (Figure \ref{fig:prob_extinct_against_L}). 

In conservation biology, we may be interested in the probability of extinction of groups of individuals rather than of a single individual's lineage. If we assume each individual's probability of extinction is independent, the probability $d_n(z)$ of an initial population of size $z$ going extinct by time $n$ is
\begin{equation}
d_n(z) = (d_n)^{z}.
\end{equation}
If the probability of extinction of a single individual's lineage is less than one, then the probability of extinction for the entire population tends asymptotically toward 0 as $z$ increases. The probability of ultimate extinction for an initial population of size $z$ can similarly be determined by
\begin{equation}
d(z) = d^{z}.
\end{equation}
For the example above, the non-spatial probability of extinction is 0.17 when $z = 1$, as determined by Eq.\ \eqref{eq:nonspatial_ult_ext}. 
As the domain length $L$ of our modified model tends to infinity, the results on the probability of extinction tend asymptotically towards those of the standard non-spatial Galton-Watson branching process. For any $L > L^*$ and an initial population greater than one, extinction is very unlikely, regardless of how much larger than $L^*$ the length of the domain is (Figure \ref{fig:prob_extinct_against_L}). Thus under the assumptions of our model, for an initial population of a few hundred or thousand individuals, the domain need only be slightly larger than $L^*$ to ensure a very low chance of extinction. Increasing the domain length far beyond $L^*$ does little to further decrease the probability of extinction. 

Alternatively, rather than determining the critical domain size for an individual and considering the population level implications, we could set a conservation goal $d$ for a given population. 
For example, let us require an ultimate extinction probability of no more than 10\%. We know that with an initial population of $z$ individuals we need $d^{z} = 0.10$. Let us suppose $z = 1000$, so we require $d \leq 0.998$. We solve Eq.\ \eqref{eq:prob_ext_disp} to obtain $F = 0.333$ so the domain needs to be large enough to retain the proportion of individuals $1-F = 0.667$. We let $\widehat{S} = 0.667$ and solve for the corresponding critical length $L^*_{\widehat{S}} = 2.726$ necessary to achieve the chosen conservation goal.

\section{Environmental stochasticity}\label{sec:Env_stoch}

We now turn our attention to the effects of a variable environment on population dynamics. Unlike in the case of demographic stochasticity, where the variability is due to random fluctuations in individual growth and survival rates, variation in environmental conditions affects all individuals similarly and simultaneously. In order to examine the effect this type of variability has on extinction probabilities, we will allow the population growth rates to fluctuate as a stationary time series  (for examples, see \cite{Athreya1971, Lande1993, Smith1969}). 

\subsection{Individual based models with environmental stochasticity}\label{sec:IBMenv} 

There are many different ways in which we could reflect the variable nature of the environment in our model. One way is to add a noise term into an otherwise deterministic framework \cite{Melbourne2008}. This can be either additive or multiplicative depending on the model formulation and, in effect, it serves to alter the average birth rate.  We choose to incorporate environmental variability directly into the birth rate of each individual, as this is straightforward within a branching process framework and also because of the implied understanding of environmental stochasticity. We take environmental variability to affect a population by changing the growth rate of every individual in the same way, by either shifting the mean and/or variance of the probability mass functions determining growth.

We introduce a sequence of iid environmental variables $\{\varsigma_n\}$, for $n = 0, 1, 2, \ldots$ where $\varsigma_n$ describes the environmental conditions in generation $n$ selected from a countable number of environmental states $\varsigma^j$, $j = 0, 1, 2, \ldots$. We denote the probability of a certain environment $\varsigma^j$ occurring for a given generation as $h_j$. Each $\varsigma^j$ uniquely determines the probability generating function of the reproductive rates of each individual which are now determined by the probability mass function $\{p_i^j\}$ comprised of the probabilities of an individual having $i$ surviving offspring, given environment $\varsigma^j$. 

To incorporate this random environmental variable into our IBM, we begin each simulation with an initial population of size $Z_0 = z$ evenly distributed throughout the domain at generation $n = 0$. We first randomly determine the environment for generation $\varsigma_n$ which corresponds to a set of reproductive probabilities described by the probability generating function $f_{\varsigma_n}$. We then determine the number of each individual's offspring according to $f_{\varsigma_n}$ and finally select the  position of the offspring after the dispersal period according to a chosen dispersal kernel. Individuals who settle within the domain survive and repeat this process over the next generation, while those who settle outside are considered lost. As with the IBM presented in Section \ref{sec:IBMs}, this is computationally costly for large populations and results for different reproductive and environmental probabilities are difficult to standardize or compare. We again rely on these results from the IBM on cumulative extinction probabilities and the critical domain length as a benchmark against which we measure and compare our spatially implicit approximations.


\subsection{Population level models with environmental stochasticity}\label{sec:pop_level_env}

We again rescale the IBM up to the population level using the Central Limit Theorem. This rescaling best mimics the IBM when the population size is not too small, due to the constraints of the Central Limit Theorem, but even for an initial population of a single individual, we find that it closely predicts the extinction probabilities. 

By the same assumptions as in Section \ref{sec:pop_level}, we approximate the dynamics at the population level by Eq.\ \eqref{eq:approximation_IBM_eqn}. Both $\bar{r}_{\bar{N}}$ and $\bar{A}_{\bar{N}}$ are random variables drawn from the same types of distributions as in Section \ref{sec:pop_level}, the difference being that we now incorporate environmental variation. As in the IBM, individual reproductive probabilities in each generation are influenced by a random variable describing the environmental conditions over that generation $\varsigma_n$. This may affect both the mean and variance of the population's per capita growth rate $\bar{r}_{\bar{N}}$. As in Section \ref{sec:pop_level}, we approximate the proportion of the population which settles in the domain $\bar{A}_{\bar{N}}$ by a normal distribution with mean $\widehat{S}$. We again require a quasi-extinction threshold ($\bar{N} < 1$) since the population will never achieve zero unless either $\bar{A}_{\bar{N}}$ or $\bar{r}_{\bar{N}}$ are zero, signifying a catastrophic collapse of the entire population in one generation.

\subsubsection{Example of population level approximation}\label{sec:pop_lev_approx_env}

\begin{figure}[!h]
	\begin{center}
		\includegraphics[width=1.00\textwidth]{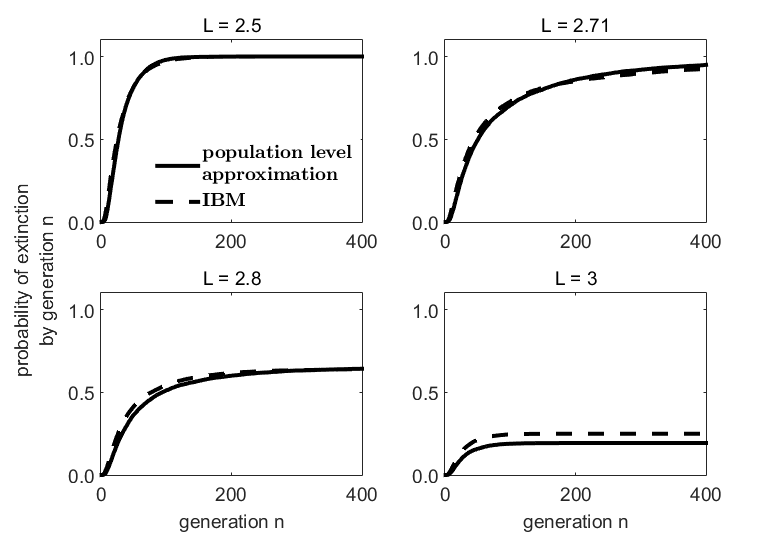}
	\caption{ As described in Section \ref{sec:pop_lev_approx_env}, the $\widehat{S}$ population level approximation to the IBM incorporating both environmental and demographic stochasticity closely approximates the cumulative extinction probabilities of the IBM. Domain lengths were chosen to be around the critical domain size $L^* = 2.703$ of the corresponding deterministic IDE \cite{Reimer2015}. This plot was generated over 10000 simulations for an initial population of 10 individuals}
	\label{fig:population_level_stochastics_environmental}
	\end{center}
\end{figure}

Assume three different environmental states $\varsigma^1$, $\varsigma^2$, and $\varsigma^3$, which occur with corresponding probabilities $h_1 = 0.4$, $h_2 = 0.4$ and $h_3 = 0.2$ respectively. Assume further that each of these environments has a corresponding expected growth rate of $\xi^1 = 1.6$, $\xi^2 = 1.5$, and $\xi^3 = 1.3$ and variance $\mbox{Var}^1 = 0.44$, $\mbox{Var}^2 = 0.45$, and $\mbox{Var}^3 = 0.61$. The expected value of $\bar{r}$ over all possible environments is 
\begin{equation}
\mathbb{E}[\bar{r}] = \xi^1 h_1 + \xi^2 h_2 + \xi^3 h_3 = 1.5.
\end{equation}
Approximating $\bar{A}$ in Eq.\ \eqref{eq:approximation_IBM_eqn} by $\widehat{S}$ for a Laplace dispersal kernel Eq.\ \eqref{eq:Laplace} yields a close prediction of the probability of extinction of the IBM (Figure \ref{fig:population_level_stochastics_environmental}).

\begin{figure}[!h]
	\begin{center}
		\includegraphics[width=0.70\textwidth]{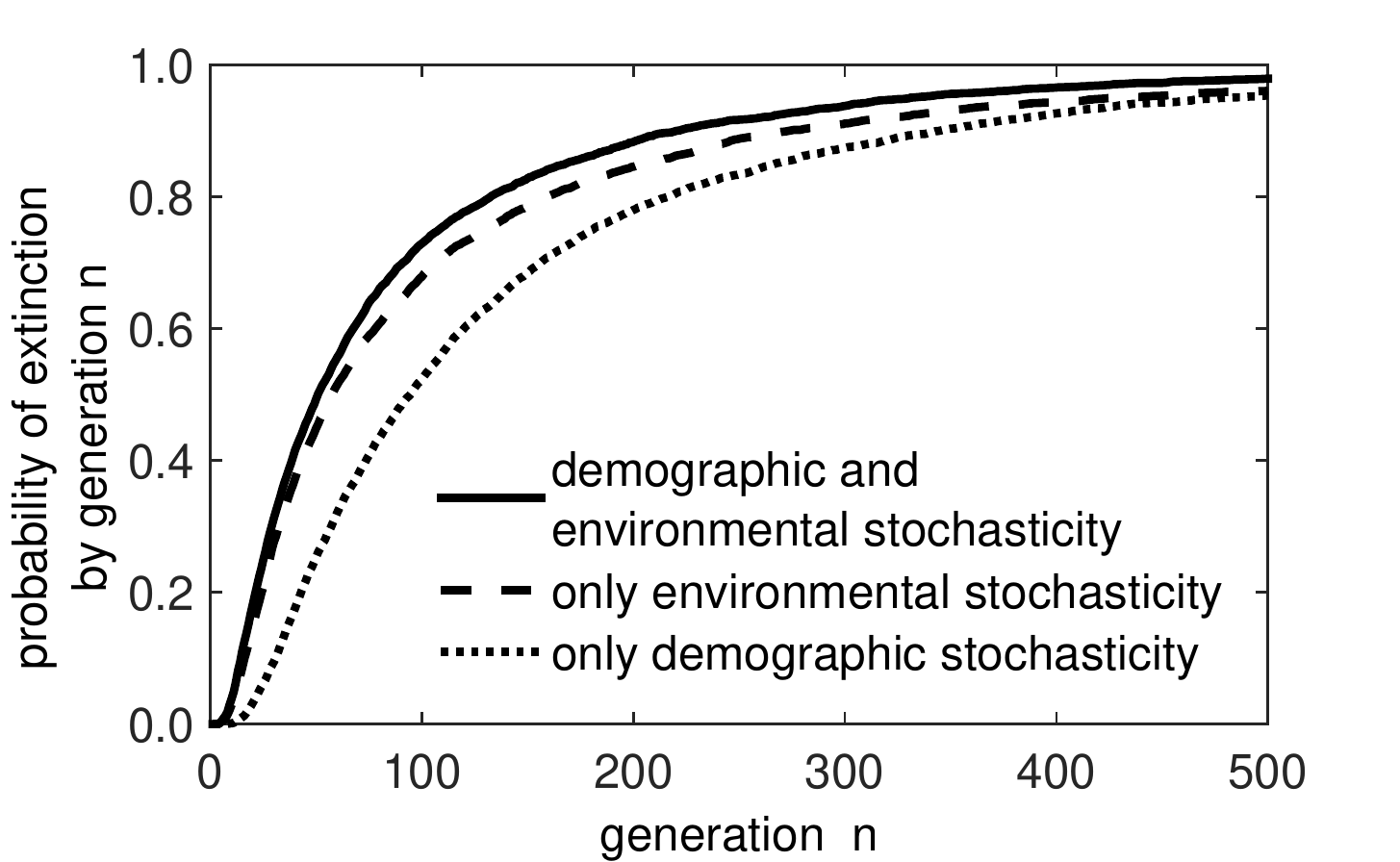}
	\caption{Probability of extinction for the $\widehat{S}$ population level approximation to the IBM under three different kinds of stochasticity as described in Section \ref{sec:pop_lev_approx_env} for a domain of length $L = 2.7$. Observe that the combination of both demographic and environmental stochasticy results in the highest probabilities of extinction. We here set the quasi-extinction threshold as $\bar{N}_n < 1$. This figure was generated over 10000 simulations for an initial population of 10 individuals}
	\label{fig:pop_level_env_vs_dem_vs_both}
	\end{center}
\end{figure}

We use this $\widehat{S}$ population level approximation to examine the difference between the probability of extinction of a population subject to both environmental and demographic stochasticity or to only one or the other. In this example, for a population subject to only demographic stochasticity, it is as though $h_2 = 1$ and $h_0 = h_3 = 0$. When the population is subject only to environmental stochasticity, $\mbox{Var}^i = 0$ for all $i$ values. Our model results agree with the intuitive notion that a population subject to multiple types of stochasticity has a higher probability of extinction (Figure \ref{fig:pop_level_env_vs_dem_vs_both}).

\subsection{A modified branching process approximation to the IBM; }\label{sec:Branching_Env} 

We again approximate the IBM by a spatially implicit branching process in order to obtain analytic results on the probability of extinction. To include environmental stochasticity, we incorporate random environments into branching processes of the Galton-Watson type, known as branching processes in random environments (BPRE). This was first done for iid random environments \cite{Smith1969} and subsequent results have been generalized to include any stationary ergodic sequence \cite{Athreya1971}. Unfortunately, as was shown by Smith and Wilkinson \cite{Smith1969} in their initial study, ``the elegant functional equations that play such a vital role in the theory of the classical Galton-Watson process, and many published generalizations thereof, do not arise in the present study'' \cite{Smith1969}. Results on the probability of extinction by a given generation tend to be approximations rather than explicit expressions of expected results. For insight into approximations for the time to extinction, or extinction probabilities of supercritical BPREs, see \cite{Agresti1975, Dekking1987, DSouza1997, Geiger2001, Grey1991, Grey1993, Wilkinson1969}.

We can, however, explicitly determine a criticality condition for the probability of ultimate extinction, where the long time behaviour changes from certain extinction to extinction with probability strictly less than one.

\subsubsection{Standard BPREs}\label{sec:BPRE}

We briefly describe the general BPRE framework of Smith and Wilkinson \cite{Smith1969} in order to be able to modify it to include implicit spatial dependence. Consider a sequence of iid random variables $\{\varsigma_n\}$, 
each selected from countably many possible environments $\{\varsigma^i\}$ according to the probability mass function $\{h_i\}$. Take the reproductive probabilities of each individual to be dependent on the environment in a given generation so that $p_i$, the probability of an individual producing $i$ offspring in one generation for a given environment $\varsigma^j$ is now $p_i(\varsigma^j) = p_i^j$. This results in a family of probability generating functions for each environment $\varsigma^j$,
\begin{equation}
f^j(s) = \sum_{\lambda=0}^{\infty}{p_\lambda^j\,s^\lambda}, \quad \quad s \in [0,1].
\end{equation}
The expected number of offspring in a fixed environment $\varsigma^j$ is then
\begin{equation}
\mathbb{E}[r^j] = \xi^j = f^j\, '(1)
\end{equation}
and $\{\xi_n\}$ is a sequence of iid random variables \cite{Smith1969}. Convention dictates that we assume $P\{\xi^j < \infty\} = 1$ for all environments as well as $P\{p_0^j + p_1^j < 1\} > 0$ 
for at least one value of $j$ (that is, the generating function is strictly convex on the unit interval for at least one environment) \cite{Wilkinson1969}. $\{Z_n\}$ is a sequence of iid random variables whose state space is the non-negative integers describing the population size in the $n^{th}$ generation given an initial population size of $Z_0 = z$. We assume $Z_0 = 1$ for the remainder of this section unless otherwise stated \cite{Smith1969}.

Smith and Wilkinson determined that certain extinction (which they termed ``mortality'') occurs for a given environmental state space if
\begin{equation}
\mathbb{E}[\ln{\xi}] \leq 0,
\end{equation}
where the expectation is taken over all possible environments \cite{Smith1969}. The case where $\mathbb{E}[\ln{\xi}] = 0$ is called ``critical'' and cases where it is strictly less than 0, ``subcritical''. ``Supercritical'' or ``immortal'' is used to describe the case when the probability of ultimate extinction is strictly less than one, and Smith and Wilkinson \cite{Smith1969} have shown that the following two conditions are both necessary and sufficient for supercriticality:
\begin{align}
&\mathbb{E}[\ln{\xi}] > 0, \quad \mbox{and} \nonumber\\
&\mathbb{E}[\,|\ln{(1-p_0)}|\,] < \infty,  \label{eq:supercrit_cond}
\end{align}
where again the expectation is over all possible environments. The first condition intuitively corresponds to the deterministic requirement that the reproductive rate be greater than one to avoid stability of the zero steady state in a map. The second condition serves to ensure that ``catastrophes'' do not occur, wiping out the entire population in one time step \cite{Smith1969}. 

\subsubsection{A spatially implicit modified BPRE}

We again incorporate the probability $F$ 
of offspring loss due to dispersal outside the domain, but now within random environments. Note that here we assume $F$ does not depend on environmental conditions. We formulate the probability generating functions incorporating $F$ by grouping our growth rates according to how many offspring successfully settle inside the domain \textit{after} the dispersal period, denoting these post-dispersal rates as $\tilde{p}_i^j$. For a given generation and environmental variable $\varsigma^j$, this results in the same $\tilde{p}_i^j$ as in Eq.\ \eqref{eq:coeff_eqn}, but now dependent on the environment $\varsigma^j$.
$\tilde{p}_i^j$ is the sum of all the possible ways in which an individual can produce $i$ offspring which survive the dispersal phase in environment $\varsigma^j$ over one generation. The probability generating function is thus
\begin{align}
f^j(s) = \tilde{p}_0^j + \tilde{p}_1^js + \tilde{p}_2^js^2 + \tilde{p}_3^js^3 + \ldots 
\end{align}
and the expected reproductive value in a given environment $\varsigma^j$ is 
\begin{align}
\xi^j &= f^j\,'(1) = \tilde{p}_1^j + 2\tilde{p}_2^j + 3\tilde{p}_3^j + \ldots \nonumber \\
	&= \sum_{q=1}^{\infty}{q\tilde{p}_q^j} = \sum_{q=1}^{\infty}{\sum_{v = q}^\infty{q\,{v \choose q,v-q}p_v^j(1-F)^qF^{v-q}}} \nonumber\\
	&= \mathbb{E}[r^j]\, (1-F),\label{eq:induction_result}
\end{align}
again by induction (Appendix \ref{Appendix:induction}). As in the case with a constant environment (Section \ref{sec:spatial_bp}), the expected reproductive rate in a given environment is the product of the expected reproductive rate of the non-spatial model and the proportion of individuals that settle within the domain. 

Applying the conditions for extinction outlined in Eq.\ \eqref{eq:supercrit_cond} for the non-spatial branching process, the criticality condition in the spatially implicit case is
\begin{equation}
\mathbb{E}[\ln \xi^j] = \mathbb{E}[\ln \mathbb{E}(r^j)] + \ln(1-F) = 0,
\label{eq:crit_cond_bpre}
\end{equation}
where the expectation is taken over all possible environments. For a given set of environmental conditions and corresponding reproductive rates, we can determine the critical proportion $F^*$ which can be lost from the domain before extinction is certain. For any $F < F^*$, the population will be supercritical and for any $F \geq F^*$, eventual extinction occurs with probability one.

\subsubsection{BPRE example}\label{sec:BPRE_example}

\begin{figure}[!h]
	\begin{center}
		\includegraphics[width=1.0\textwidth]{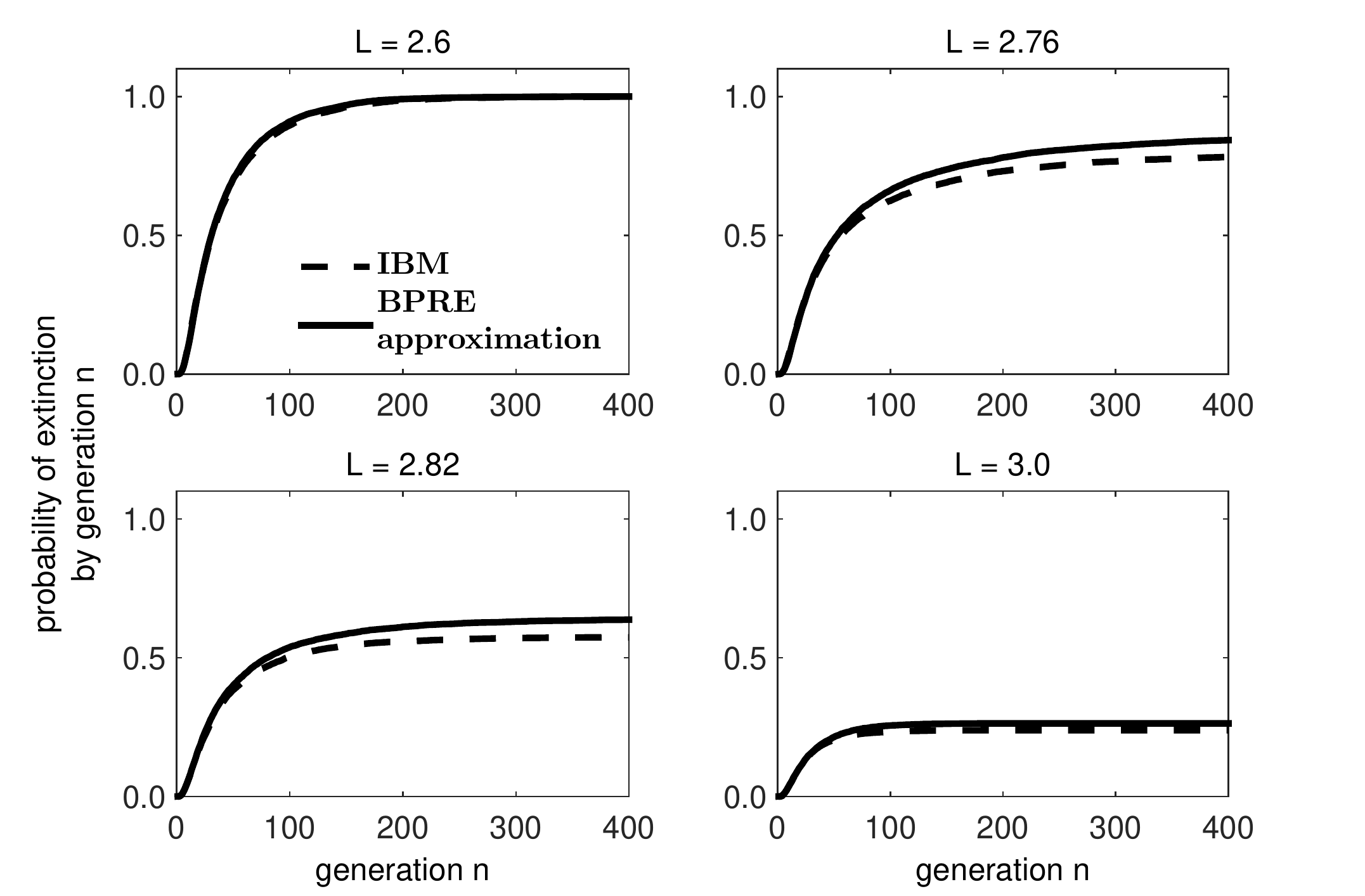}
	\caption{BPRE closely approximates the IBM with demographic and environmental variability, using $\widehat{S}$ to approximate the proportion of successful settlers dispersing according to a Laplace kernel. Initial population size is $Z_0 = 10$ and 10000 simulations were used to obtain these results. Models are as described in Section \ref{sec:BPRE_example} and domain lengths were chosen around the critical lengths of 
	$L^* = 2.703$ and $L^*_{\widehat{S}} = 2.744$}
	\label{fig:stochastic_IBM_environmental}
	\end{center}
\end{figure}

We build on our previous examples and assume an environmental state space of three possible environments, $\varsigma^1$, $\varsigma^2$, and $\varsigma^3$ which occur with probability $h_1 = 0.4$, $h_2 = 0.4$ and $h_3 = 0.2$. Each of these environments has a corresponding probability mass function for the random variable $r$ which describes the number of offspring of an individual. We assume that for the first environment, reproductive probabilities are $\{p_0^1 = 0.1, p_1^1 = 0.2, p_2^1 = 0.7\}$, for the second they are $\{p_0^2 = 0.1, p_1^2 = 0.3, p_2^2 = 0.6\}$, and for the third, $\{p_0^3 = 0.2, p_1^3 = 0.3, p_2^3 = 0.5\}$. This results in mean reproductive values $\xi^1 = 1.6$, $\xi^2 = 1.5$, and $\xi^3 = 1.3$ and respective variances $\mbox{Var}^1 = 0.44$, $\mbox{Var}^2 = 0.45$, and $\mbox{Var}^3 = 0.61$. The expected number of offspring $\xi$ over all environments is
\begin{equation}
\xi = \mathbb{E}[r] = \xi^1 h_1 + \xi^2 h_2 + \xi^3 h_3 = 1.5.
\end{equation}
Ignoring loss via dispersal,  
\begin{equation}
\mathbb{E}[\ln f\,'(1)] = 0.402 > 0
\label{eq:nonspatial}
\end{equation}
and so the non-spatial process is supercritical. To determine the criticality condition while implicitly including space, we incorporate loss of individuals outside the domain with probability $F$ into our probabilities $p_i^j$, so that the probability generating function $f^j(s)$ now has spatially implicit coefficients as in Eq.\ \eqref{eq:coeff_eqn}. 
%
%
For our three environments, the above probabilities, and the knowledge that $\xi^j = \mathbb{E}\left[r^j\right](1-F) = (p_1^j + 2p_2^j)(1-F)$ from Eq.\ \eqref{eq:induction_result}, we have 
\begin{align}
\xi^1 & = 1.6\, (1-F)\nonumber\\
\xi^2 & = 1.5\, (1-F) \label{eq:induction_result2}\\
\xi^3 & = 1.3\, (1-F),\nonumber 
\end{align}
and so for the probability mass function with values $h_1, h_2, h_3$ as above, 
\begin{align}
\mathbb{E}[\ln \xi] &= h_1\ln(1.6(1-F)) + h_2 \ln(1.5(1-F)) + h_3\ln(1.3(1-F)) \nonumber \\
      & = 0.402 + \ln(1-F).
\end{align}
As expected, this is the sum of the non-spatial expected reproductive rate from Eq.\ \eqref{eq:nonspatial} and $\ln(1-F)$. We find the critical proportion $F^*$ from Eq.\ \eqref{eq:crit_cond_bpre} by solving
\begin{equation}
\mathbb{E}[\ln \xi] = 0.402 + \ln(1-F) = 0, 
\end{equation} 
which results in $F^* = 0.331$. We use $\widehat{S}$ to approximate the proportion of individuals successfully settling within the domain, so $\widehat{S} = 1-F$, and then solve for the corresponding domain lengths using Eq.\ \eqref{eq:hatS} to obtain the branching process approximation to the critical domain size for an ultimate extinction probability strictly less than one. If we assume dispersal according to a Laplace kernel Eq.\ \eqref{eq:Laplace}, $L_{\widehat{S}}^* = 2.744$ (Figure \ref{fig:stochastic_IBM_environmental}). Again, from \cite{Reimer2015}, the critical domain size of the stochastic model is larger than that of the corresponding deterministic IDE model with an intrinsic reproductive value of $1.5$, logistic growth and a Laplace dispersal kernel resulting in a critical domain size of $L^* = 2.703$. 

\section{Discussion}

The critical domain size deemed necessary for population persistence is highly dependent on model structure and assumptions. We here have considered populations which would typically be modelled using an IDE framework, with discrete pelagic and dispersal periods. We were interested in ways to include stochasticity - both demographic and environmental - into estimators of critical domain size for these populations. 

First considering only demographic stochasticity, we created an IBM in order to simulate explicitly each individual's variable reproductive rate and also its random dispersal distance. While the IBM most closely mimics the assumptions of the deterministic IDE framework, it does not easily lend itself to comparison between different kernels or parameters. It is also slow to simulate for large populations due to the computational costs associated with having each individual disperse independently. We thus developed two approximations to the IBM which both perform well for a variety of domain lengths and initial population sizes. Using the Central Limit Theorem, we found that scaling the IBM up to the population level results in a close approximation to the probability of extinction and provides faster simulation results. To obtain analytic results, we modified a Galton-Watson branching process to implicitly include space. This also closely approximated the IBM while allowing for closer examination of the affects of per capita growth rates and dispersal distances on population decline. IBMs rarely yield analytic insight and this spatially implicit branching process approximation allows us to scale individual variability up to obtain results at the population level. 

Following the same model progression but now also including environmental stochasticity, we modified our IBM so that each individual reproduced and dispersed independently but now with the probability distributions of reproduction influenced by the environment for a given generation. We again used the Central Limit Theorem to obtain population level simulation results on the probability of extinction for various domain lengths. We then modified the branching process to include random environments using a BPRE framework. Unlike when we only considered demographic variability, existing results on BPREs do not allow us to analytically determine the long time probability of extinction or the probability of extinction by a given generation. Rather we must rely on the criticality condition which separates the cases where extinction is certain from those where the population goes extinct with a probability less than one. 

Both the population level, as well as the branching process models, relied on $\widehat{S}$ to approximate the proportion of individuals successfully retained following each dispersal event. As was shown for the deterministic case in \cite{Reimer2015}, this provides evidence that the details of the dispersal kernel are not important, but rather it is the proportion of the population retained in the domain which determines population persistence.  

Regardless of whether we are considering the results from the IBM, the population level approximations, or the branching process models, the critical domain size of the stochastic models was always larger than that of the corresponding deterministic IDE models. This implies that the critical domain size of the deterministic model is not large enough to sustain a population subject to either demographic or environmental stochasticity. This disparity between the stochastic and deterministic models increased for both decreasing growth rates and increasing variance in vital rates. As was shown by Watson and Galton (1875), certain extinction is determined only by the expected reproductive rate, however, if extinction is not certain, then the variance will affect the ultimate extinction probability. The probability of extinction by a given generation thus relies on both the expected reproductive rate and the variance in the probability mass function $\{p_i\}$. The higher the expected reproductive rate, the lower the probability of extinction, and the greater the variance, the higher the probability of extinction.

From studying the branching process models, we concluded that once the domain is larger than the criticality condition dictates, any additional area added to the domain does not greatly affect the probability of extinction for large initial populations. This result followed from the assumption of independence of all individuals, which may not be true in real populations. Density dependent reproduction either at low or high densities may occur due to resource limitation or the effect of density on mating probabilities. In addition, when including environmental stochasticity we did not allow for the possibility that dispersal behaviour may be dependent on the environmental conditions for a given generation. Including environmental dependence of dispersal behaviour may add additional realism. Further extensions to these models could also include patchy domains connected via dispersal, as well as including stage or age structure into the populations, as has been done for deterministic models using structured population models \cite{Caswell2006}.

Because both the population level and branching process approximations rely on $\widehat{S}$ rather than on a comprehensive understanding of the dispersal kernel, it would be just as easy to have both the population level approximations and the branching process models represent a domain in two or three dimensions, rather than along a one-dimensional domain. Given empirical data, $\widehat{S}$ could be parametrized in this way, rather than requiring an explicit form for $k(x,y)$ which may be beneficial in systems where the mechanistic underpinnings of dispersal are unknown. For our modified BPRE, it is sufficient if experimental information can be obtained on both reproductive rates in a range of environments and the proportion of individuals retained within the area of interest.

In spite of the fact that analytic results on the critical domain size of the IBM do not exist, we were able to use branching processes to address the questions of critical domain size for populations subject to demographic and environmental stochasticity. While the population level approximations provided faster simulation results and insight into the efficacy of using the $\widehat{S}$ approximation to dispersal success in this framework, it is the branching process approximations which allowed for insight into the affects of varying demographic rates, dispersal distances, and environments. Demographic and environmental variability are known to be important in determining the population persistence and these tools may aid in understanding their comparative effects on the effect of a finite domain on extinction risk. 

\section*{Acknowledgments}
The authors acknowledge the support of the Rhodes Trust and the Natural Sciences and Engineering Research Council of Canada.

\newpage

\appendix

\section{Proof of Eq.\ \eqref{eq:ind_result} and Eq.\ \eqref{eq:induction_result}} \label{Appendix:induction}

Here we prove 
\begin{align}
\xi &= \sum_{q=1}^{\infty}{q\bar{p}_j} = \sum_{q=1}^{\infty}{\sum_{v = q}^\infty{q\,{v \choose q,v-q}p_v(1-F)^qF^{v-q}}} \nonumber\\
	&= \mathbb{E}[r]\, (1-F) \label{eq:conjecture}
\end{align} 
from Eq.\ \eqref{eq:ind_result} using induction. This proof also holds for Eq.\ \eqref{eq:induction_result}, with a change of notation to represent the dependence of $\xi$ on the environment $\zeta^j$. We show that for all $k \in \mathbb{N}$, 
\begin{align}
\xi &= \sum_{q=1}^{k}{\sum_{v = q}^{k}{q\,{v \choose q,v-q}p_v(1-F)^qF^{v-q}}}\nonumber\\
	&= \mathbb{E}[r]\, (1-F). 
\end{align}
First, let $k = 1$ (i.e. a parent either has zero or one offspring in a generation). Then 
\begin{align}
\xi &= p_1(1-F) \nonumber\\
&= \mathbb{E}[r]\, (1-F).
\end{align}
Next, assume that Eq.\ \eqref{eq:conjecture} holds for any $k \in \mathbb{N}$, say
\begin{align}
\xi_k &= \sum_{q=1}^{k}{\sum_{v = q}^{k}{q\,{v \choose q,v-q}p_v(1-F)^qF^{v-q}}}\nonumber \\
		&= \mathbb{E}[r]_k\, (1-F).
\end{align}
Therefore,
\begin{align}
\xi_{k+1} &= \sum_{q=1}^{k+1}{\sum_{v = q}^{k+1}{q\,{v \choose q,v-q}p_v(1-F)^qF^{v-q}}} \nonumber\\
	&= \xi_k + \underbrace{\sum_{q=1}^{k+1}{q\,{k+1 \choose q,(k+1)-q}p_{k+1}(1-F)^qF^{(k+1)-q}}}_{(a)}.
\end{align} 
We now show that
\begin{equation}
(a) = (k+1)p_{k+1}(1-F)
\label{eq:second_conjecture}
\end{equation}
for all $k$. This problem can be simplified to the proof that 
\begin{equation}
\sum_{q=1}^{k}{\frac{(k-1)!}{(q-1)!(k-q)!}(1-F)^{q-1}F^{k-q}} = 1.
\end{equation}
If we here let $p = q-1$, and $n = k-1$, this becomes
\begin{equation}
\sum_{p = 0}^{n}{\frac{n!}{p!(n-p)!}(1-F)^pF^{n-p}} 
\end{equation}
which is the binomial formula for $\left[(1-F) + F\right]^n$ which is 1. So now from Eq.\ \eqref{eq:conjecture}, we have that 
	\begin{align}
		\xi_{k+1} &= \xi_k + (k+1)p_{k+1}(1-F)\nonumber\\
			&= \mathbb{E}[r]_k (1-F) + (k+1)p_{k+1}(1-F)\\
			&= \mathbb{E}[r]_{k+1}(1-F)\quad \square\nonumber
	\end{align}

\bibliography{arXivdoc}
\bibliographystyle{plain}

\end{document}